\documentclass[epj]{webofc}
\usepackage[varg]{txfonts}   
\usepackage{graphicx,color} 
\usepackage{bm}       
\usepackage{amsmath}  
\usepackage{amssymb}
\woctitle{21st International Conference on Few-Body Problems in Physics}
\begin{document}
\title{Recent results on the nucleon resonance spectrum and structure from the CLAS detector}
\author{V.~I.~Mokeev\inst{1,2}\fnsep\thanks{\email{mokeev@jlab.org}},  
        I.~G.~Aznauryan\inst{3}, 
        V.~D.~Burkert \inst{1}, \and
        R.~W.~Gothe \inst{4}
\,\,  for the CLAS Collaboration }

\institute{Thomas Jefferson National Accelerator Facility, Newport News, Virginia 23606, USA 
\and
           Skobeltsyn Nuclear Physics Institute and Physics Department at Moscow State University, Moscow 119899, Lenninskie gory, Russia
\and
           Yerevan Physics Institute, 375036, Yerevan, Armenia
\and
           University of South Carolina, Columbia, South Carolina 29208, USA	   
          }

\abstract{The CLAS detector at Jefferson Lab has provided the dominant part of all available worldwide data on exclusive meson electroproduction off protons in the resonance region. New results on the $\gamma_{v}pN^*$ transition amplitudes (electrocouplings) are available from analyses of the CLAS data and will be presented. Their impact on understanding of hadron structure will be discussed emphasizing the credible access to the dressed quark mass function that has been achieved for the first time by a combined analysis of the experimental results on the electromagnetic nucleon elastic and  $N \rightarrow N^*$ transition form factors. We will also discuss further convincing evidences for a new baryon state  $N^{\, '}(1720)3/2^+$ found in a combined analysis of charged double pion photo- and electroproduction cross sections off the protons.}  

\maketitle
\section{Introduction}
\label{intro}

Studies of the excited nucleon state ($N^*$) spectrum and structure from results on resonance electroexcition amplitudes ($\gamma_{v}pN^*$ electrocouplings) at different photon virtualities ($Q^2$) represent an important direction in the broad efforts on the exploration of the non-perturbative strong interaction and its emergence from QCD \cite{Az13,Cr14}. They are the only source of information on many facets of the non-perturbative strong interaction in the generation of $N^*$-states with different quantum numbers as bound systems of an infinite amount of the QCD current quarks and gauge gluons \cite{Cr14}. For the first time $\gamma_{v}pN^*$ electrocouplings of most well established nucleon resonances with masses below 1.8 GeV have been extracted from the data on exclusive meson electroproduction off protons measured with CLAS at Jefferson Lab at photon virtualities up to 5.0 GeV$^2$ \cite{Bu12,Mo14,Park15}. In these proceedings we will present new results on $\gamma_{v}pN^*$ electrocouplings from CLAS and will discuss their impact on our understanding of the $N^*$ structure, and in particular, the opportunity to map-out the dressed quark mass function which encodes the emergence of hadron masses and quark-gluon confinement from QCD. Furthermore, we will present new results from a combined analysis of the charged double pion photo- and electroproduction off protons that strongly support the existence of a new baryon state $N^{\, '}(1720)3/2^+$.

\section{Evaluation of the resonance parameters from the CLAS data}
\label{sec-1}

Nucleon resonance electroexcitations are described by two transverse ($A_{1/2}(Q^2)$, $A_{3/2}(Q^2)$) and one longitudinal ($S_{1/2}(Q^2)$) electrocoupling amplitudes. These electrocouplings offer access to the resonance structure. The CLAS detector has contributed the lion's share of the world data on all essential exclusive meson electroproduction channels in the resonance excitation region including $N\pi$, $\eta p$, $KY$, and $\pi^+\pi^-p$ electroproduction off protons with nearly complete coverage of the final hadron phase space \cite{Bu12}. The observables measured with the CLAS detector are stored in the CLAS Physics Data Base \cite{db15}. The available in the resonance excitation region observables are listed in Table~\ref{tab-1}. 
 
\begin{table}[htb!]
\begin{center}
\caption{\label{tab-1} Observables for exclusive meson electroproduction off protons that have been measured with the CLAS detector in the resonance excitation region and stored in the CLAS Physics Data Base \cite{db15}: CM-angular distributions for the final mesons ($\frac{d\sigma}{d\Omega}$); beam, target, and beam-target asymmetries ($A_{LT'}$, $A_{t}$, $A_{et}$); and recoil hyperon polarizations ($P^{'}$, $P^{0}$).}
\begin{tabular}{|c|c|c|c|}
\hline
Hadronic       &  $W$-range    & $Q^2$-range     & Measured observables \\
final state    &  GeV          & GeV$^2$         &     \\ 
\hline
$\pi^+ n$      &  1.10-1.38     & 0.16-0.36      & $\frac{d\sigma}{d\Omega}$ \\
               &  1.10-1.55     & 0.30-0.60      & $\frac{d\sigma}{d\Omega}$ \\
               &  1.10-1.70     & 1.70-4.50      & $\frac{d\sigma}{d\Omega}$, $A_{LT'}$ \\
               &  1.60-2.00     & 1.80-4.50      &  $\frac{d\sigma}{d\Omega}$    \\
\hline	       
$\pi^0 p$      &  1.10-1.38     & 0.16-0.36      & $\frac{d\sigma}{d\Omega}$ \\
               &  1.10-1.68     & 0.40-1.15      & $\frac{d\sigma}{d\Omega}$, $A_{LT'}$, $A_{t}$, $A_{et}$ \\
               &  1.10-1.39     & 3.00-6.00      & $\frac{d\sigma}{d\Omega}$  \\
\hline     
$\eta p$       &  1.50-2.30     & 0.20-3.10      & $\frac{d\sigma}{d\Omega}$ \\
\hline     
$K^{+}\Lambda$ &  1.62-2.60     & 1.40-3.90      & $\frac{d\sigma}{d\Omega}$ \\
               &  1.62-2.60     & 0.70-5.40      & $P^{'}$, $P^{0}$ \\
\hline     
$K^{+}\Sigma^0$ &  1.62-2.60     & 1.40-3.90      & $\frac{d\sigma}{d\Omega}$ \\
                &  1.62-2.60     & 0.70-5.40     & $P^{'}$ \\
\hline     
$\pi^+\pi^-p$   &  1.30-1.60     & 0.20-0.60      & Nine single-differential \\
                &  1.40-2.10     & 0.50-1.50      & cross sections \\
\hline
\end{tabular}
\end{center}
\end{table}

So far, most of the results on $\gamma_{v}pN^*$ electrocouplings have been extracted from independent analyses of $\pi^+n$, $\pi^0p$, and $\pi^+\pi^-p$ exclusive electroproduction data off the protons. A total of nearly 160,000 data points (d.p.) on unpolarized differential cross sections, longitudinally polarized beam asymmetries, and longitudinal target and beam-target asymmetries for $N\pi$ electroproduction off protons were obtained with the CLAS detector at $W <$ 2.0 GeV and 0.2 GeV$^2 < Q^2 <$ 6.0 GeV$^2$. The data have been analyzed within the framework of two conceptually different approaches: a unitary isobar model (UIM) and dispersion relations (DR) \cite{Az09,Park15}. The UIM describes the $N\pi$ electroproduction amplitudes as a superposition of $N^*$ electroexcitations in the $s$-channel, non-resonant Born terms and $\rho$- and $\omega$- t-channel contributions. The latter are reggeized, which allows for a better description of the data in the second- and third-resonance regions. The final-state interactions are treated as $\pi N$ rescattering in the K-matrix approximation~\cite{Az09}. In the DR approach, dispersion relations relate the real to the imaginary parts of the invariant amplitudes that describe the $N\pi$ electroproduction. Both approaches provide good and consistent description of the $N\pi$ data in the range of $W$ $<$ 1.7 GeV and $Q^2$ $<$ 5.0 GeV$^2$, resulting in $\chi^2$/d.p. $< 2.9$. 
  
The $\pi^+\pi^- p$ electroproduction data from CLAS \cite{Ri03,Fe09} provide for the first time information on nine independent single-differential and fully-integrated cross sections binned in $W$ and $Q^2$ in the mass range $W <$ 2.0 GeV and at photon virtualities of 0.25 GeV$^2 < Q^2 < 1.5$ GeV$^2$. The analysis of the data allowed us to develop the JM reaction model~\cite{Mo09,Mo12} with the goal of extracting resonance electrocouplings as well as $\pi\Delta$ and $\rho p$ hadronic decay widths. This model incorporates all relevant reaction mechanisms in the $\pi^+\pi^-p$ final-state channel that contribute significantly to the measured electroproduction cross sections off protons in the resonance region, including the $\pi^-\Delta^{++}$, $\pi^+\Delta^0$, $\rho^0 p$, $\pi^+N(1520)\frac{3}{2}^-$, $\pi^+N(1685)\frac{5}{2}^+$, and $\pi^-\Delta(1620)\frac{3}{2}^+$ meson-baryon channels as well as the direct production of the $\pi^+\pi^-p$ final state without formation of intermediate unstable hadrons. The contributions from well established N$^*$ states in the mass range up to 2.0 GeV were included into the amplitudes of $\pi\Delta$ and $\rho p$ meson-baryon channels by employing a unitarized version of the Breit-Wigner ansatz \cite{Mo12}. The current analysis of the preliminary $\pi^+\pi^-p$ photoproduction data from CLAS at 1.6 GeV $<$ $W$ $<$ 2.0 GeV \cite{Mo15} revealed the presence of a three-body final state interaction that is parameterized and fit to the nine single-differential cross sections. The JM model provides a good description of $\pi^+\pi^- p$ differential cross sections at $W$ $<$ 1.8 GeV and 0.2 GeV$^2 < Q^2 < 1.5$ GeV$^2$ with $\chi^2$/d.p. $< 3.0$, and the preliminary $\pi^+\pi^- p$ photoproduction data are also described successfully at $W$ up to 2.0 GeV~\cite{Mo15}. The achieved quality of the CLAS data description allows us to isolate the resonant contributions and to determine both resonance electrocouplings and $\pi N$, $\pi \Delta$, and $\rho N$ decay widths fitting them to the measured observables.

Resonance electrocouplings have been obtained from various CLAS data in the exclusive channels: $\pi^+n$ and $\pi^0p$ at $Q^2 < 5.0$ GeV$^2$ in the mass range up to 1.7 GeV, $\eta p$ at $Q^2 < 4.0$ GeV$^2$ in the mass range up to 1.6 GeV, and $\pi^+\pi^-p$ at $Q^2 < 1.5$ GeV$^2$ in the mass range up to 1.8 GeV \cite{Bu12,Mo14}. For the first time photocouplings and $\pi\Delta$ and $\rho N$ hadronic decay widths of all well established resonances in the mass range from 1.6 GeV to 2.0 GeV that decay preferentially to the N$\pi\pi$ final states have become available in the analysis of the preliminary CLAS $\pi^+\pi^-p$ photoproduction data \cite{Mo15}. The resonance electrocouplings obtained from these $\pi^+\pi^-p$ photoproduction data are consistent with the published RPP results~\cite{rpp} from analyses of $N\pi$ photoproduction, confirming the reliability of the updated JM model in extracting resonance parameters. 

\section{Highlights of the recent results on the $N^*$ spectrum and structure from the CLAS data}
\label{sec-2}
\subsection{Impact of the new CLAS results on studies of the $N^*$ structure}
\label{ssec-2}

The studies of the $N(1440)1/2^+$ and $N(1520)3/2^-$ resonances with the CLAS detector~\cite{Az09,Mo12} have provided the dominant part of the worldwide available information on their electrocouplings in a wide range of photon virtualities 0.25 GeV$^2 < Q^2 < 5.0$ GeV$^2$. Currently the $N(1440)1/2^+$ and $N(1520)3/2^-$ states, together with the $\Delta(1232)3/2^+$ and $N(1535)1/2^-$ resonances~\cite{Bu12}, represent the most explored excited nucleon states. Furthermore, results on the $\gamma_{v}pN^*$ electrocouplings for the high-lying $N(1675)5/2^-$, $N(1680)5/2^+$, and $N(1710)1/2^+$ resonances have recently been determined for the first time from the CLAS $N\pi$ data at 1.5 GeV$^2 < Q^2 < 4.5$ GeV$^2$ \cite{Park15}. 

\begin{figure}[htb!]
\centering
\includegraphics[width=4.6cm,clip]{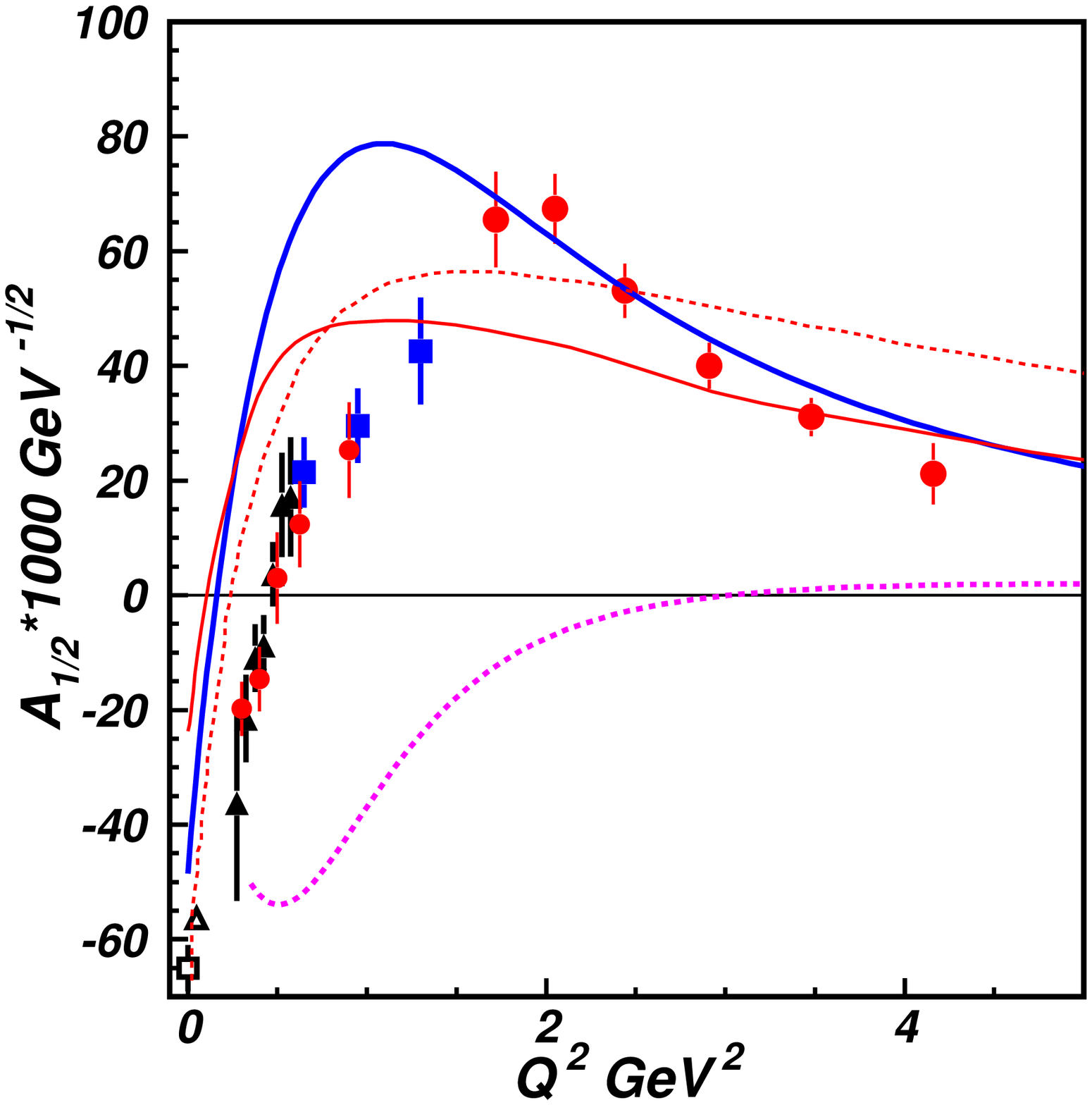}
\includegraphics[width=4.6cm,clip]{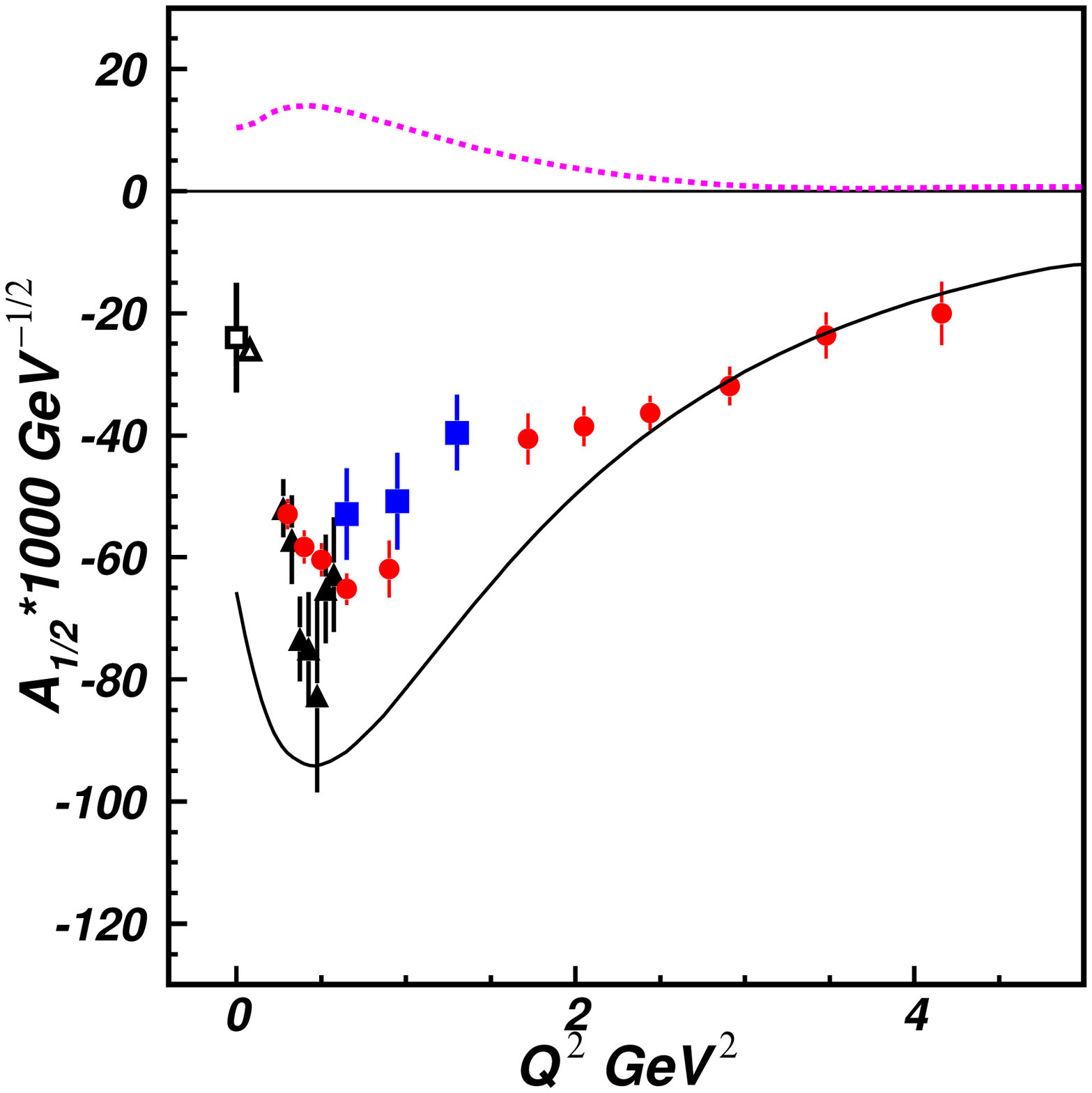}
\includegraphics[width=4.6cm,clip]{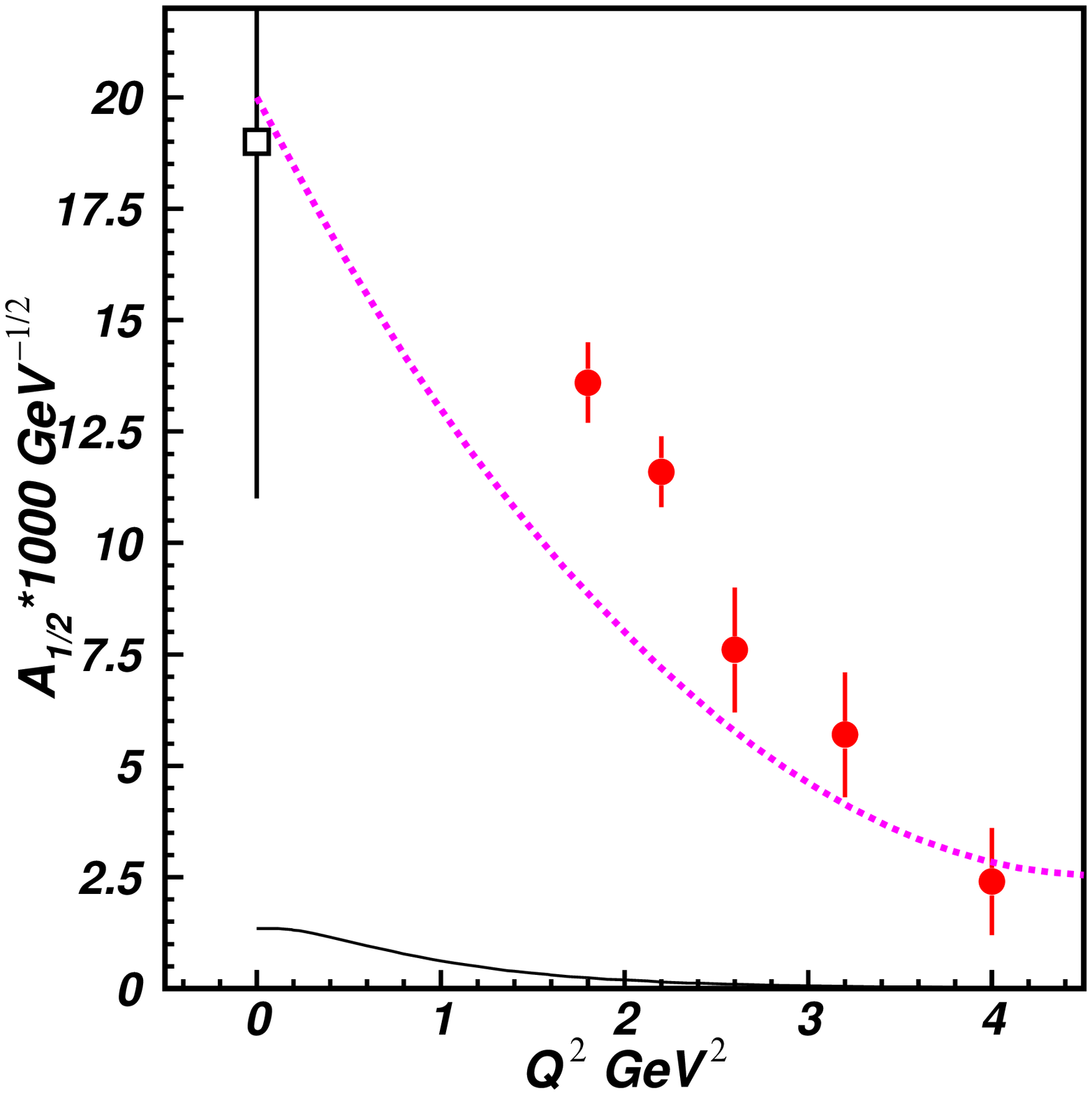} 
\caption{(Color Online) $A_{1/2}$ $\gamma_vpN^*$ electrocouplings of the $N(1440)1/2^+$ (left), $N(1520)3/2^-$ (center), and  $N(1675)5/2^-$ (right) resonances from analyses of the CLAS electroproduction data off protons in the $N\pi$ - \cite{Az09,Park15} (red circles) and $\pi^+\pi^-p$ channels~\cite{Mo12} (black triangles) with new preliminary results from $\pi^+\pi^-p$ channel~\cite{Mo14} (blue squares). The electrocoupling results are shown in comparison with the DSEQCD - \cite{Cr15a} (blue thick solid) and constituent quark model calculations~\cite{Az12} (thin red solid), ~\cite{Ob14} (thin red dashed), and \cite{Sa12} (thin black solid). The meson-baryon cloud contributions are presented by the magenta thick dashed lines. In case of the $N(1440)1/2^+$ resonance (left) they are based on the calculated DSEQCD results and the extracted electrocoupling data, whereas the absolute values at the resonance poles taken from Argonne-Osaka coupled channel analysis \cite{Lee08} are shown for $N(1520)3/2^-$ (center) and  $N(1675)5/2^-$ (right). Photocouplings are taken from RPP \cite{rpp} (black open squares) and the CLAS data analysis~\cite{Dug09} of $N\pi$ photoproduction.}
\label{p11d13d15}       
\end{figure}

\begin{figure}[htp!]
\begin{center}
\includegraphics[width=4.9cm]{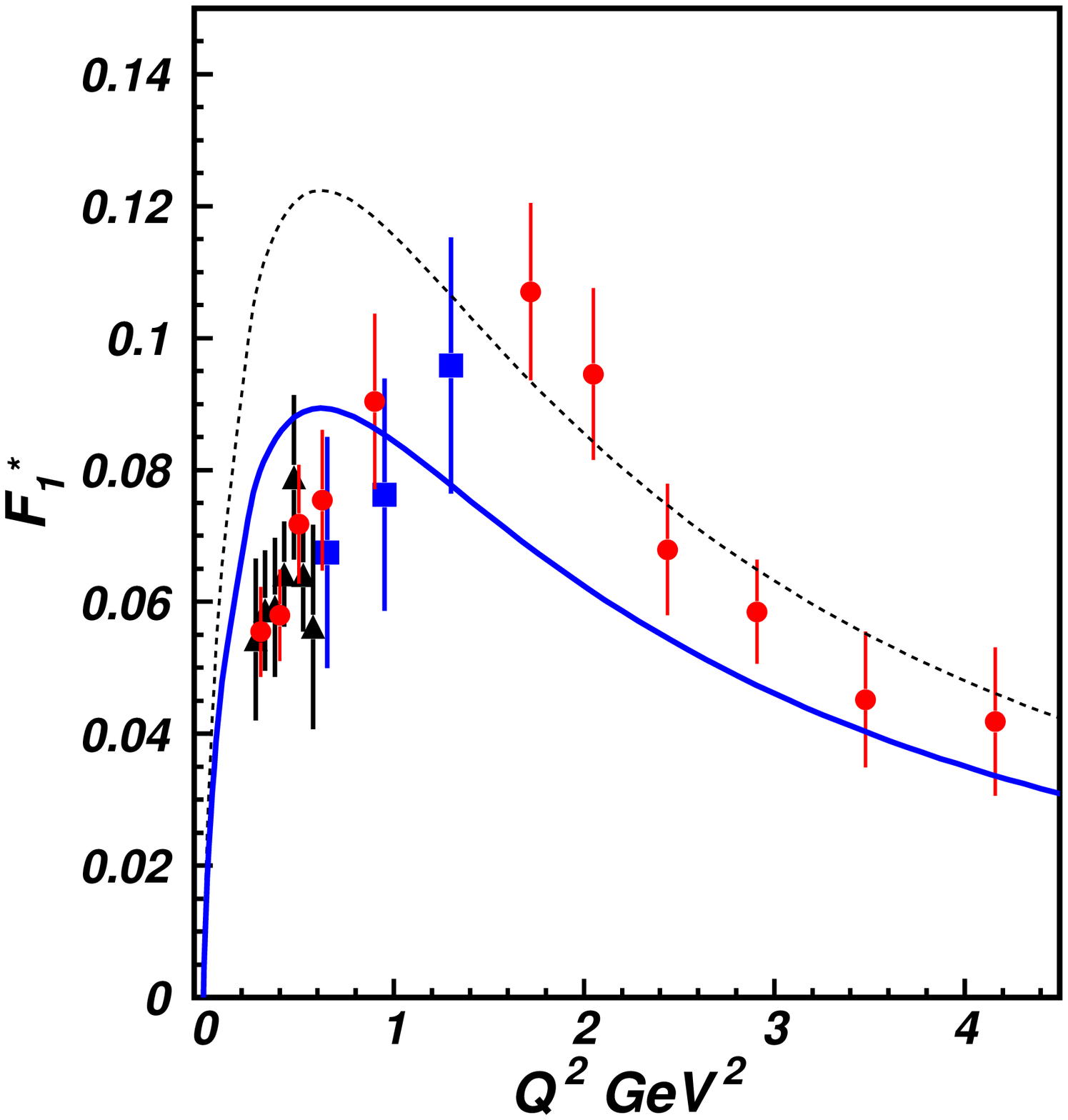}
\includegraphics[width=4.9cm]{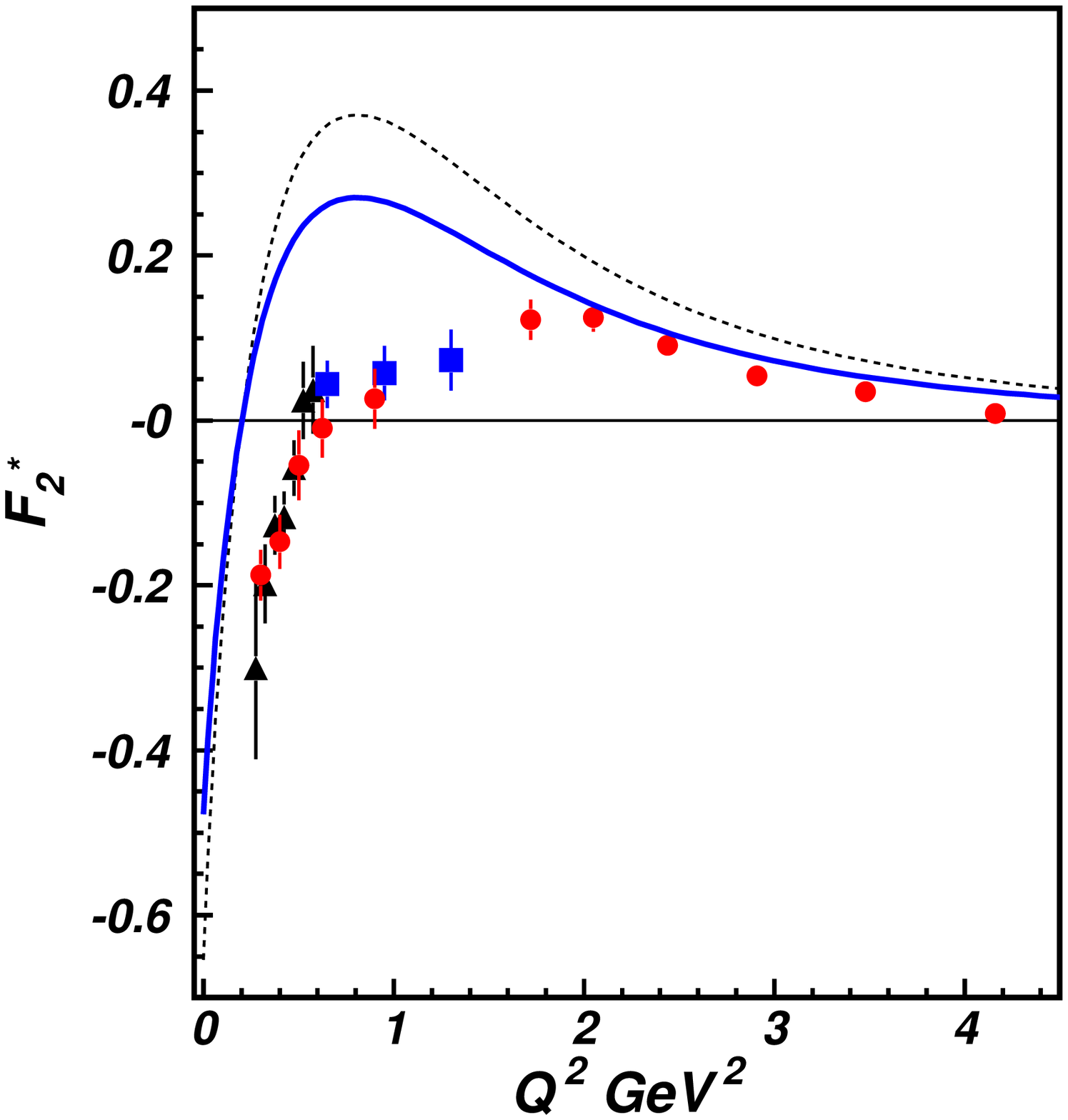}
\vspace{-0.1cm}
\caption{(Color Online) The $F_1^*$ and $F_2^*$ $p \to N(1440)1/2^+$ electromagnetic transition form factors. Experimental results from analyses of the CLAS electroproduction data off protons on $N\pi$~\cite{Az09} (red circles) and on $\pi^+\pi^-p$~\cite{Mo12} (black triangles) and new preliminary results from $\pi^+\pi^-p$ channel (blue squares). These data are shown in comparison with DSEQCD evaluations~\cite{Cr15a} that start from the QCD Lagrangian and incorporate the contributions from dressed quarks only (black dashed line) and the dressed quark contributions after accounting for the admixture from meson-baryon cloud as described in Section~\ref{sec-2} (blue solid line).}
\label{p11datdse}
\end{center}
\end{figure} 

Figure~\ref{p11d13d15} shows the mentioned electrocoupling results for the $N(1440)1/2^+$, $N(1520)3/2^-$, and  $N(1675)5/2^-$ resonances together with the preliminary results on the $N(1440)1/2^+$ and $N(1520)3/2^-$ electrocouplings from the CLAS $\pi^+\pi^-p$ electroproduction off protons at 0.5 GeV$^2 < Q^2 < 1.5$ GeV$^2$~\cite{Mo14}. The $N \rightarrow N(1440)1/2^+$ Dirac ($F_{1}^*$) and Pauli ($F_{2}^*$) electromagnetic transition form factors, as computed from these data, are shown in Fig.~\ref{p11datdse}. Consistent results for the $\gamma_vpN^*$ electrocouplings of the $N(1440)1/2^+$ and $N(1520)3/2^-$ resonances, that have been determined in independent analyses of the dominant meson electroproduction channels, $N\pi$ and $\pi^+\pi^-p$, demonstrate that the extraction of these fundamental quantities is reliable, since good data description is achieved in the major electroproduction channels, that have quite different background contributions. This consistency also strongly suggests that the reaction models, described in  Sec.~\ref{sec-1}, provide the basis for reliable evaluations of the $\gamma_{v}pN^*$ electrocouplings, and it is therefore possible to determine these electrocouplings for the majority of $N^*$ resonances that decay preferentially to the  $N\pi$ and$/$or $N\pi\pi$ final states. 

Due to the rapid progress in the field of DSEQCD (Dyson-Schwinger Equations of QCD) studies of excited nucleon states~\cite{Cr14,Cr15}, the first evaluations of the electromagnetic transition $p \to N(1440)1/2^+$ Dirac $F_1^*$ and Pauli $F_2^*$ form factors starting from the QCD Lagrangian have recently become available~\cite{Cr15a}. They are shown in Fig.~\ref{p11datdse} by the black dashed lines and were computed accounting only for the contributions from three dressed quarks. However, we have to account for the fraction of meson-baryon degrees of freedom in the wave functions of the ground and excited nucleon states. We choose to estimate this by multiplying the $p \to N(1440)1/2^+$ electromagnetic transition form factors computed within the DSEQCD approach~\cite{Cr15a} by a common $Q^2$-independent factor that is fit to the data for $Q^2 > 3.0$~GeV$^2$, where the quark core is expected to be the biggest contributor. The $p \to N(1440)1/2^+$ electromagnetic transition form factors obtained in this manner are shown in Fig.~\ref{p11datdse} by the solid blue lines. They offer a good description of the experimental results for $Q^2 > 1.5$~GeV$^2$. Discrepancies at smaller $Q^2$ are related to meson-baryon-cloud contributions, which are beyond the scope of the current DSEQCD approach \cite{Cr13,Cr15,Cr15a}. The dressed quark mass function used in the DSEQCD computations of the $p \to N(1440)1/2^+$ electromagnetic transition form factors~\cite{Cr15a} is {\it exactly the same} as the one employed in the previous evaluations of the electromagnetic nucleon form factors - and the $p \to \Delta$ transition form factors~\cite{Cr15,Cr13}. This success strongly supports: a) the relevance of dynamical dressed quarks, with properties predicted by the DSEQCD approach~\cite{Cr14}, as constituents of the quark core in the structure of the ground and excited nucleon states; and b) the capability of the DSEQCD approach~\cite{Cr15,Cr15a} to map out the dressed quark mass function from the experimental results on the $Q^2$-evolution of the nucleon elastic - and $p \to N^*$ electromagnetic transition form factors or rather $\gamma_vpN^*$ electrocouplings. Studies of the dressed quark mass function will address the most challenging and still open problems of the Standard Model on the nature of the dominant part of the hadron mass, quark-gluon confinement, its emergence from QCD, and its relation to dynamical chiral symmetry breaking which is expected to be a source of more than 98\% of the hadron mass in universe~\cite{Cr14}.

Recent advances in the development of constituent quark models make it possible to extend the $Q^2$ range in comparison with DSEQCD approaches where a reasonable description of the $\gamma_vpN^*$ electrocouplings is achieved by taking both the contributions, the quark core and meson-baryon cloud, into account. The two models~\cite{Az12,Ob14}, that account for the meson-baryon cloud contributions, allowed us to considerably improve the description of $N(1440)1/2^+$ electrocouplings at $Q^2 < 2.0$~GeV$^2$, confirming the relevance of the meson-baryon degrees of freedom for the $N(1440)1/2^+$ structure at these distances, as it is shown in Fig.~\ref{p11d13d15} (left). The credible DSEQCD evaluation of the quark core contributions to the electrocouplings of the $N(1440)1/2^+$ allows us to evaluate the meson-baryon cloud contributions as the difference between the fit of the experimental data at higher $Q^2$ and the quark core electroexcitation amplitudes from DSEQCD~\cite{Cr15a}. The  meson-baryon cloud contributions extracted in this manner are shown in Fig.~\ref{p11d13d15} (left). The CLAS results on the $\gamma_vpN^*$ electrocouplings of the $N(1520)3/2^-$ resonance are shown in Fig.~\ref{p11d13d15} (center). The quark core contributions to these electrocoupling results have been explored within the framework of the two conceptually different approaches: a) the hypercentral constituent quark model~\cite{Sa12}, and b) the Bethe-Salpeter approach that employs structureless constituent quarks with momentum-independent quark mass and an instanton quark-quark interaction~\cite{Met12}. Both approaches provide a reasonable description of the experimental results at $Q^2 > 1.0$~GeV$^2$ as shown in Fig.~\ref{p11d13d15} (center). At smaller photon virtualities they are unable to reproduce the CLAS experimental results. The absolute values of the meson-baryon cloud contributions from Argonne-Osaka analysis \cite{Lee08}, also shown in Fig.~\ref{p11d13d15} (center), are maximal at small photon virtualities, where the discrepancies between the quark model expectations and the experimental data are largest. The absolute value of the meson-baryon dressing amplitudes for the $A_{1/2}$ electrocouplings of the $N(1520)3/2^-$ are much smaller than those for the $S_{1/2}$ and $A_{3/2}$ electrocouplings as well as for the $A_{1/2}$ electrocoupling of the $N(1440)1/2^+$. This makes the $A_{1/2}$ electrocoupling of the $N(1520)3/2^-$ particularly attractive for the studies of quark degrees of freedom within the structure of the $N(1520)3/2^-$ resonance. On the other hand, the results on the electrocouplings of the $N(1675)5/2^-$ resonance are already reasonably described by only accounting for the meson-baryon cloud contributions from Argonne-Osaka approach~\cite{Lee08}, see Fig.~\ref{p11d13d15} (right), since the quark core electroexcitation amplitudes vanish in the limit of the contribution from only the leading SU(6)-spin-flavor configuration and remain almost negligible accounting for the quark configuration mixing, see black solid line in Fig.~\ref{p11d13d15} (right) \cite{Sa12}. These results offer for the first time an almost direct access to meson-baryon cloud contributions to the $N^*$ structure \cite{Bu15} because of strong suppression for the quark core electroexcitations. 

Analyses of the CLAS results strongly suggest that the structure of nucleon resonances for $Q^2 < 5.0$~GeV$^2$ is determined by a complex interplay between the inner core of three dressed quarks bound into a system with the quantum numbers of the nucleon resonance and the external meson-baryon cloud which also depends substantial on the quantum numbers of the excited nucleon state. The studies of resonance electrocouplings over the full spectrum of excited nucleon states of different quantum numbers are hence critical in order to explore different components in the $N^*$ structure.

\subsection{Further evidence for the existence  of a new $N^{\, '}$(1720)3/2$^{+}$ baryon state}
\label{ssec-3}

The combined analysis of $\pi^+\pi^-p$ electroproduction and preliminary photoproduction data from CLAS provides further evidence for the existence of a new N$^{\, '}$(1720)3/2$^{+}$ baryon state. For the first time signals from the $N^{\, '}(1720)3/2^+$ candidate state have been observed in analyses of the CLAS  $\gamma_{v}p \rightarrow \pi^+\pi^-p$ exclusive electroproduction data \cite{Ri03}, which shows a pronounced structure in the $W$-dependence of fully integrated cross sections at $W$ $\approx$ 1.7 GeV and in all $Q^2$ bins covered by the measurements.  Analyses of a very limited amount of the single-differential cross sections (three of nine available) carried out in \cite{Ri03} within the framework of the oversimplified initial version of the JM model \cite{Ri20,Mo01} suggested two possible ways to describe the structure observed at W $\approx$ 1.7 GeV: a) by a dominating decay of the $N(1720)3/2^+$ resonance to the  $\pi \Delta$ final state accounting only for contributions from conventional resonances, or b) by implementing a new $N^{\, '}(1720)3/2^+$ candidate state with parameters derived from the CLAS data fit, in which case the hadronic decay widths of all known resonances remain inside the ranges reported in the PDG03 \cite{rpp03}.

Further studies of the CLAS $\gamma_{v}p \rightarrow \pi^+\pi^-p$ electroproduction cross sections \cite{Ri03} have been carried out within the framework of the current version of the JM model \cite{Mo09,Mo12,Mo14} outlined in Sec.~\ref{sec-1}. All nine single-differential cross sections were included in the data fit. In the fits we varied simultaneously electrocouplings, hadronic decay widths to the $\pi \Delta$ and $\rho N$ final states for all resonances that contribute to the region of the structure at $W \approx$ 1.7 GeV.  We also simultaneously varied the non-resonant parameters of the JM model. The hadronic decay width of each resonance remains the same in all Q$^2$-bins for the electroproduction data. The preliminary CLAS photoproduction data \cite{Mo15} and the CLAS electroproduction data \cite{Ri03} were fit independently, allowing us to examine the consistency of the resonance hadronic parameters extracted from the photo- and electroproduction data independently.  

Two fits of the nine single-differential $\gamma_{r,v}p \rightarrow \pi^+\pi^-p$ cross sections have been carried out: a) assuming only the contribution from conventional resonances (fit A), or b) a $N^{\, '}(1720)3/2^+$ candidate state was implemented in addition to the contributions from the conventional resonances (fit B). Both fit A and fit B provide a good description of the CLAS $\gamma_{r,v}p \rightarrow \pi^+\pi^-p$ photo- and electroproduction cross sections at $W$ from 1.61 to 1.81 GeV with the $\chi^2$/$d.p.$ less than 3.03 and less than 2.80, respectively. A reliable description of the resonant content implies a valuable restriction on the resonance hadronic decay widths inferred at different values of photon virtuality Q$^2$. The hadronic decay widths of each resonance should be Q$^2$-independent. The hadronic decays widths of the conventional $N(1720)3/2^+$ resonance obtained from fit A of the charged double pion photoproduction data \cite{Mo15} 
are listed in the Table~\ref{hadrp13phel} in comparison with those inferred in fit A of the   electroproduction cross sections \cite{Ri03}. 

\begin{table}[htb!]
\begin{center}
\caption{\label{hadrp13phel} N(1720)3/2$^+$ hadronic decays determined from the independent fits to the data on charged double pion photo- \cite{Mo15} and electroproduction \cite{Ri03} off protons accounting only for contributions from conventional resonances.}
\begin{tabular}{|c|c|c|c|}
\hline
 Resonance            & N$^*$ total width       & Branching fraction            & Branching fraction         \\
states                & MeV                     &  for decays to $\pi\Delta$    &  for decays to $\rho N$    \\
\hline
 N(1720)3/2$^+$       &                         &                               &                            \\
 electroproduction    &  126.0 $\pm$ 14.0       &    64\% - 100\%               & $<$ 5\%                    \\
 photoproduction      &  160.0 $\pm$ 65.0       &   14\% - 60\%                 &  19\% - 69\%               \\
\hline
\end{tabular}
\end{center}
\end{table}

The branching fractions for the $N(1720)3/2^+$ decays into the $\rho N$ final state inferred by fit A of the data on charged double pion photo- and electroproduction off protons differ by more than a factor of four. Moreover, the branching fractions for the  $N(1720)3/2^+$ decays into the $\pi \Delta$ final states obtained fitting the photo- and electroproduction data are also rather different. This makes it impossible to describe both the charged double pion photo- and electroproduction cross sections when only contributions from conventional resonances are taken into account. By implementing a new $N^{'}(1720)3/2^+$ baryon state , a successful description of all nine single-differential $\gamma_{r,v}p \rightarrow \pi^+\pi^-p$  photo- and electroproduction cross sections has been achieved. Furthermore, the hadronic decay widths of all resonances in the third resonance region as inferred from the fits at different $Q^2$ remain $Q^2$-independent in the entire range of photon virtualities up to 1.5 GeV$^2$ that is covered by the CLAS measurements \cite{Mo15,Ri03} (Table~\ref{hadr_phot_miss}.) 

\begin{table}[htb!]
\begin{center}
\caption{\label{hadr_phot_miss} Hadronic decays into the $\pi \Delta$ and $\rho N$ final states of the resonances in the third resonance region with major decays to the N$\pi\pi$ final states determined from the fits to the data on charged double pion photo- \cite{Mo15} and 
electroproduction \cite{Ri03} implementing a new $N^{ '}(1720)3/2^+$ baryon state.}
\begin{tabular}{|c|c|c|c|}
\hline
 Resonance            & N$^*$ total width         & Branching fraction            & Branching fraction         \\
states                & MeV                       & for decays to $\pi\Delta$     &  for decays to $\rho N$    \\
\hline
 $\Delta(1700)3/2^-$  &                           &                               &                           \\
 electroproduction    &  288.0 $\pm$ 14.0         &    77\% - 95\%                &   3\% - 5\%                      \\
 photoproduction      &  298.0 $\pm$ 12.0         &   78\% - 93\%                 &  3\% - 6\%                    \\
\hline
 N(1720)3/2$^+$       &                           &                               &                           \\
 electroproduction    &  116.0 $\pm$ 7.0          &    39\% - 55\%                &   23\% - 49\%                     \\
 photoproduction      &  112.0 $\pm$ 8.0          &   38\% - 53\%                 &  31\% - 46\%                    \\
\hline
N$^{\, '}$(1720)3/2$^+$ &                          &                               &                           \\
 electroproduction    &  119.0 $\pm$ 6.0         &     47\% - 64\%               &    3\% - 10\%                     \\
 photoproduction      &  120.0 $\pm$ 6.0         &     46\% - 62\%               &    4\% - 13\%                    \\
\hline
\end{tabular}
\end{center}
\end{table}

\begin{figure}[hbt!]
\begin{center}
\includegraphics[width=4.6cm,clip]{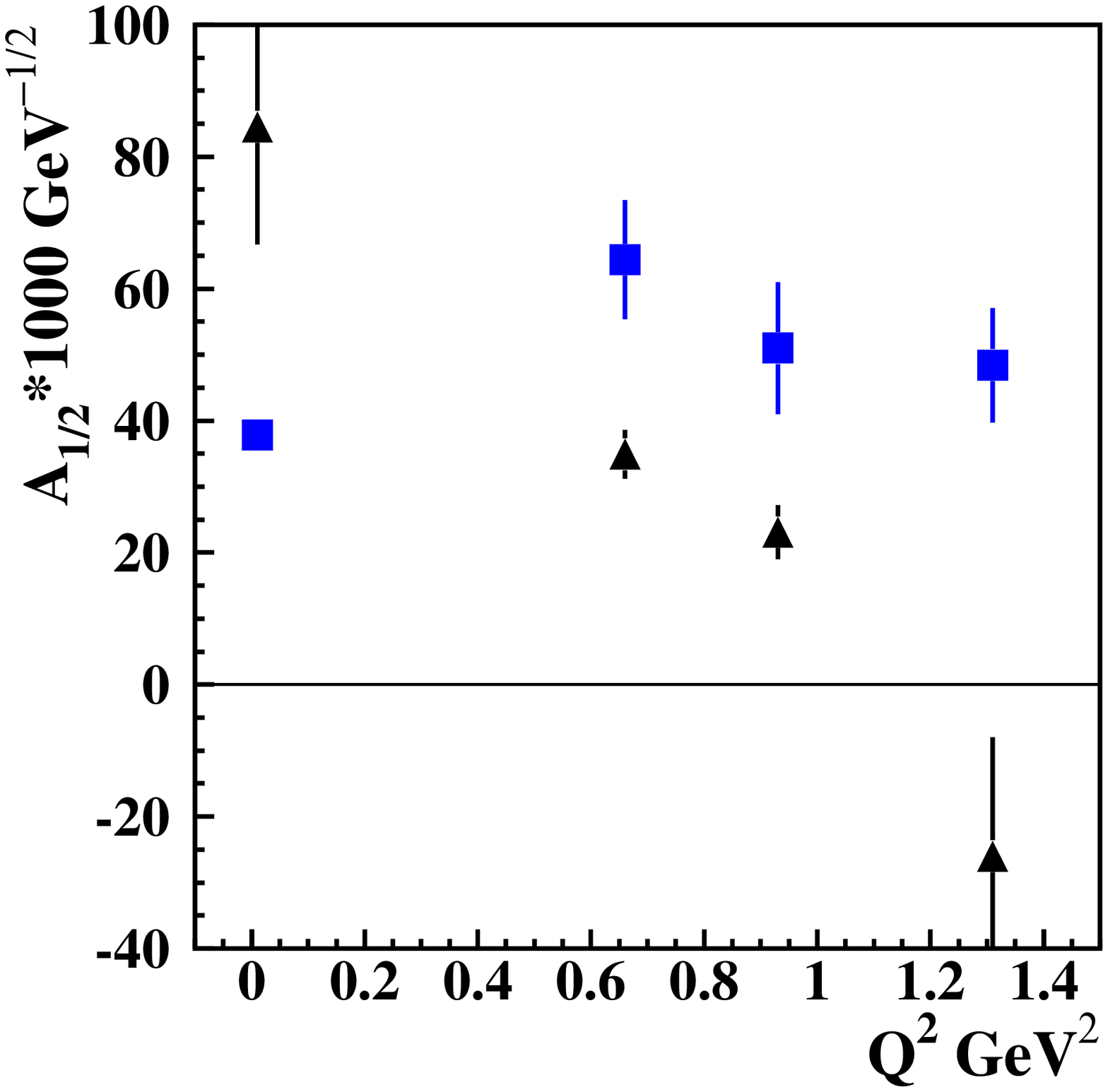}
\includegraphics[width=4.6cm,clip]{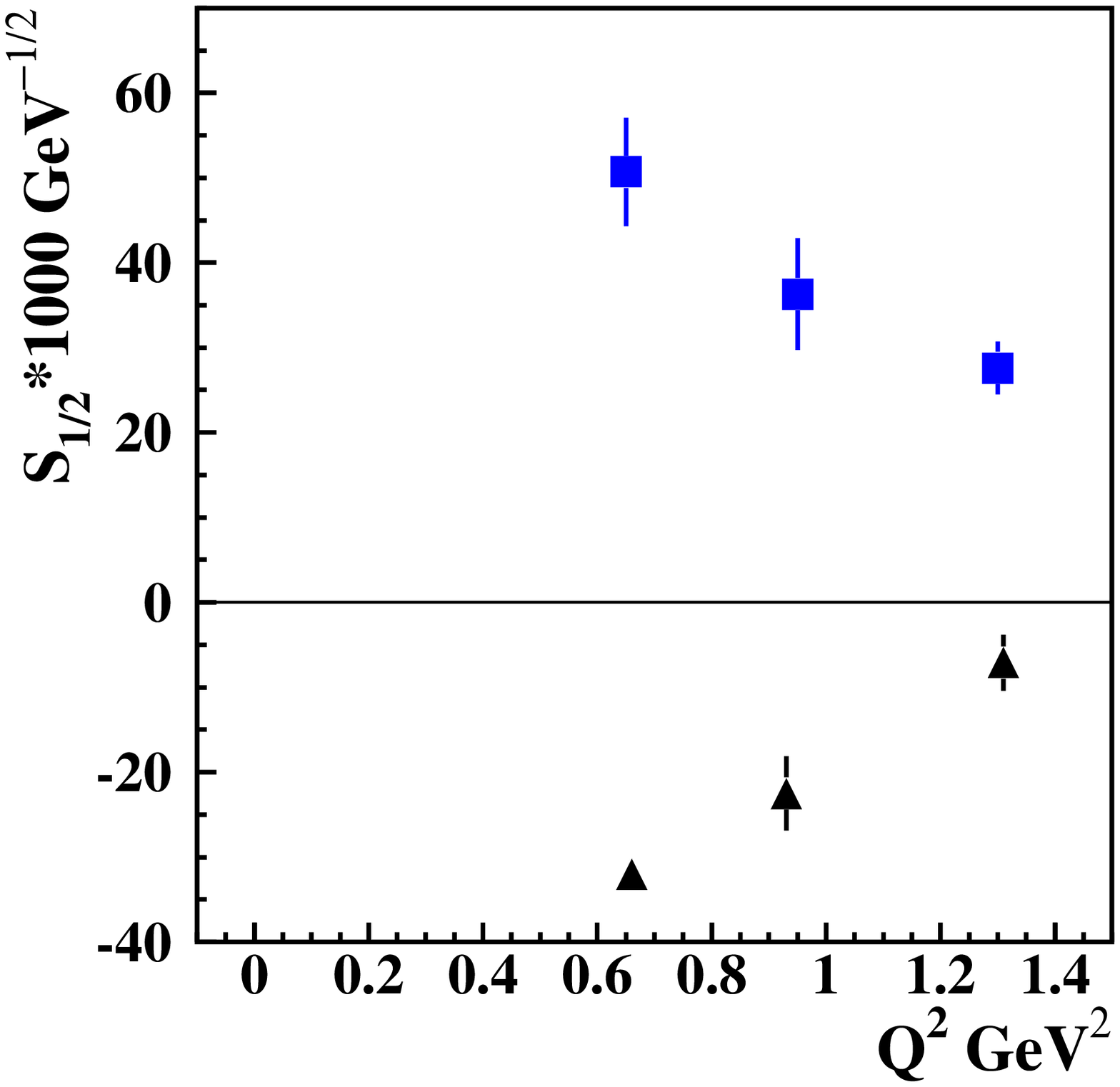}
\includegraphics[width=4.6cm,clip]{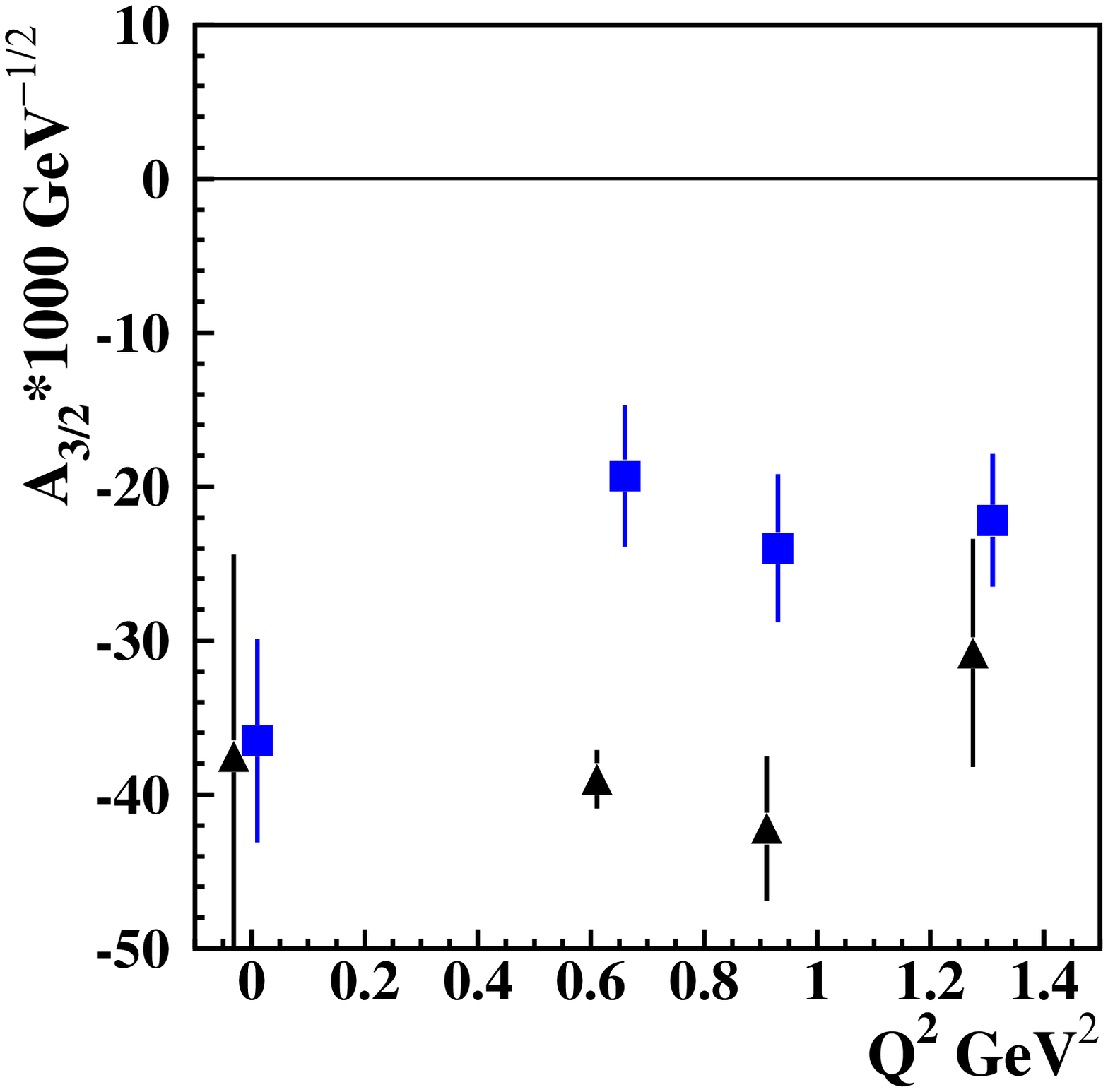}
\caption{\small (Color online) Photo- and electrocouplings of the new $N^{\, '}(1720)3/2^+$ (blue squares) and conventional $N(1720)3/2^+$ (black triangle) states as determined from the analysis of the CLAS charged double pion photo- and electroproduction data off protons~\cite{Mo15,Ri03} within the framework of JM15 model \cite{Mo09,Mo12,Mo14} outlined in Sec.~\ref{sec-1}. } \label{newconv}
\end{center}
\end{figure}

The implementation of a new $N^{\, '}(1720)3/2^+$ baryon state represents the only way that allows us to achieve a good description of the CLAS data on charged double pion photo- and electroproduction in all four $Q^2$ bins covered by the CLAS measurements, centered at 0 GeV$^2$, 0.65 GeV$^2$, 0.95 GeV$^2$, and 1.30 GeV$^2$, with $Q^2$-independent hadronic decay widths of each relevant resonance. This success provides sound evidence for the first observation of the $N^{'}(1720)3/2^+$ baryon state in the CLAS data \cite{Mo15,Ri03}. The electrocouplings of the two close-lying resonances, the new $N^{\, '}(1720)3/2^+$ and the conventional
$N(1720)3/2^+$, are superimposed in Fig.~\ref{newconv}. 

The analysis of the CLAS data within the framework of the recent JM model version \cite{Mo09,Mo12} revealed the contribution of a conventional $N(1720)3/2^+$ and a new $N^{\,'}(1720)3/2^+$ baryon state to the $\pi^+\pi^-p$ photo- and electroproduction cross sections at $W$ $\approx$ 1.7 GeV. The two states have close masses (1.743 GeV to 1.753 GeV for $N(1720)3/2^+$ and 1.715 GeV to 1.735 GeV for $N^{\,  '}(1720)3/2^+$) and the same spin-parities 3/2$^+$, but their hadronic decay widths of the $\pi \Delta$, $\rho N$ final states and $Q^2$-evolution of their electrocouplings (Fig.~\ref{newconv}) are distinctively different. All this further strengthens the claim of a new $N^{\, '}(1720)3/2^+$ baryon state found in the CLAS charged double pion photo- and electroproduction data.

\begin{acknowledgement}
We would like to acknowledge the outstanding efforts of the staff of the Accelerator and the Physics Divisions at Jefferson Lab. We are grateful to I.C. Cl\"{o}et, M.M. Giannini, C.D. Roberts, E. Santopinto, J. Segovia for their theoretical support. This work was supported in part by the U.S. Department of Energy, the National Science Foundation, the Skobeltsyn Institute of Nuclear Physics, the Physics Department at Moscow State University, the University of South Carolina, and Yerevan Physics Institute. The Southeastern Universities Research Association (SURA) operates the Thomas Jefferson National Accelerator Facility for the United States Department of Energy under the contract DE-AC05-84ER40150.
\end{acknowledgement}

%

\begin{thebibliography}{}
%
\bibitem{Az13} I.~G.~Aznauryan et al., 
Int. J. Mod. Phys. $\textbf{E22}$, 1330015
(2013).

\bibitem{Cr14} I.~C.~Cl\"{o}et and C.~D.~Roberts, 
Prog. Part. Nucl. Phys. $\textbf{77}$, 1
(2014).

\bibitem{Bu12} I.~G.~Aznauryan and V.~D.~Burkert, 
Prog. Part. Nucl. Phys. $\bf{67}$, 1
(2012).

\bibitem{Az09} I. G. Aznauryan et al., CLAS Collaboration,
Phys. Rev. C $\bf{80}$, 055203 (2009).

\bibitem{Park15} K. Park et al.,
CLAS Collaboration, 
Phys. Rev. C $\bf{91}$, 045203 (2015).

\bibitem{Mo12} V. I. Mokeev et al., CLAS Collaboration,
Phys. Rev. C $\bf{86}$, 055203 (2012).

\bibitem{Mo14} V.~I. Mokeev and I. G. Aznauryan,  
Int. J. Mod. Phys. Conf. Ser. $\bf{26}$, 146080
(2014).

\bibitem{db15} CLAS Physics Data Base, http://clas.sinp.msu.ru/cgi-bin/jlab/db.cgi.

\bibitem{Ri03} M.~Ripani et al., CLAS Collaboration, Phys. Rev. Lett. $\bf
91$, 022002 (2003).

\bibitem{Fe09} G.~V. Fedotov et al., CLAS Collaboration, 
Phys. Rev. C $\bf 79$, 015204 (2009).

\bibitem{Mo09} V.~I.~Mokeev et al., Phys. Rev. C $\bf 80$, 045212 (2009).

\bibitem{Mo15} V.I.Mokeev, Invited talk at The International Workshop on Partial
 Wave Analysis for Hadron Spectroscopy, PWA 8 / ATHOS 3, http://physics.columbian.gwu.edu/program-and-contributions.

\bibitem{rpp} J. Beringer et al., Review of Particle Physics, Phys. Rev. D $\bf{86}$, 1 (2012).

\bibitem{Dug09}  M. Dugger et al., CLAS Collaboration
Phys. Rev. C $\bf 79$, 065206 (2009).

\bibitem{Cr15} 
J. Segovia {\it et al.}, Few Body Syst. {\bf 55}, 1185 (2014).

\bibitem{Cr15a} 
J. Segovia {\it et al.}, arXiv:1504.04386 [nucl-th].

\bibitem{Cr13} 
I.C. Cl\"{o}et, C.D. Roberts and A.W. Thomas, Phys. Rev. Lett. {\bf 111}, 101803 (2013).

\bibitem{Az12}  
I.G. Aznauryan and V.D. Burkert, Phys. Rev. C {\bf 85}, 055202 (2012).

\bibitem{Ob14}  
I.T. Obukhovsky {\it et al.}, Phys. Rev. D {\bf 89}, 0142032 (2014).

\bibitem{Sa12}  
E. Santopinto and M.M. Giannini, Phys. Rev. C {\bf 86}, 065202 (2012).

\bibitem{Met12} 
M. Ronninger and B. Ch. Metsch, Eur. Phys J. A {\bf 49}, 8 (2012).

\bibitem{Lee08} 
B. Julia-Diaz {\it et al.}, Phys. Rev. C {\bf 77}, 045205 (2008).

\bibitem{Bu15} 
I.G. Aznauryan and V.D Burkert, Phys. Rev. C {\bf 92}, 015203 (2015).

\bibitem{Ri20} M. Ripani et al.,
Nucl. Phys. A $\bf{672}$, 220 (2000).

\bibitem{Mo01} V. I. Mokeev et al.,
Phys. Atom. Nucl. $\bf{64}$, 6 (2001).

\bibitem{rpp03} D. E. Groom et al.,
Eur. Phys, J. C $\bf{15}$, 1 (2000).

\end{thebibliography}
%
%

\end{document}